\documentclass[twocolumn,superscriptaddress,aps,prl]{revtex4-1}
\usepackage{amsmath}
\usepackage{graphicx}
\usepackage{dcolumn}
\usepackage{color}
\usepackage{ulem}

\begin{document}

\title{Many-Body Echo}
\author{Yang-Yang Chen}
\affiliation{Shenzhen Institute for Quantum Science and Engineering,
and Department of Physics, Southern University of Science and
Technology, Shenzhen 518055, China}
\affiliation{Center for Quantum Computing, Peng Cheng
Laboratory, Shenzhen 518055, China}
\affiliation{CAS Key Laboratory of Quantum Information, University of Science and Technology of China, Hefei 230026, China}
\author{Pengfei Zhang}
\affiliation{Institute for Advanced Study, Tsinghua University, Beijing,
100084, China}
\author{Wei Zheng}
\affiliation{Department of Physics, The University of Hong Kong, Hong Kong, China}
\author{Zhigang Wu}
\email{wuzg@sustech.edu.cn}
\affiliation{Shenzhen Institute for Quantum Science and Engineering,
and Department of Physics, Southern University of Science and
Technology, Shenzhen 518055, China}
\affiliation{Center for Quantum Computing, Peng Cheng
Laboratory, Shenzhen 518055, China}
\author{Hui Zhai}
\email{hzhai@tsinghua.edu.cn}
\affiliation{Institute for Advanced Study, Tsinghua University, Beijing,
100084, China}
\affiliation{Center for Quantum Computing, Peng Cheng
Laboratory, Shenzhen 518055, China}
\date{\today}

\begin{abstract}
In this letter we propose a protocol to reverse a quantum many-body dynamical
process. We name it  ``many-body echo" because the underlying physics
is closely related to the spin echo effect in nuclear magnetic resonance systems. We consider a periodical modulation of the interaction strength in a weakly interacting Bose condensate, which resonantly excites quasi-particles from the condensate. A dramatic phenomenon is that, after pausing the
interaction modulation for half a period and then continuing on with the same
 modulation, nearly all the excited quasi-particles in the resonance
modes will be absorbed back into the condensate. During the intermediate half period, the free evolution introduces a $\pi$ phase, which plays a role reminiscent
of that played by the $\pi$-pulse in the spin echo. Comparing our protocol with another one
 implemented by the Chicago group in a recent experiment, we find
that ours is more effective at reversing the many-body process. The
difference between these two schemes manifests the physical effect of the
micro-motion in the Floquet theory. Our scheme can be generalized to other
periodically driven many-body systems.
\end{abstract}

\maketitle

How to reverse a quantum many-body dynamical process is a question of great
interest, especially in recent discussions of quantum many-body chaos and quantum
information scrambling~\cite{chaos1,chaos2,chaos3}. Ultracold atomic gases provide a unique
platform to address this kind of questions because of the following two reasons.
Firstly, unlike other artificial quantum systems such as nuclear magnetic resonance (NMR)
and trapped ions, where the number of qubits is currently limited to below a few
hundreds, ultracold atomic gases are many-body systems containing a macroscopically large number of
quantum particles. Secondly, in contrast to electronic
systems in condensed matter materials where phonons are inevitably present
and will cause decoherence and dissipation, ultracold atomic gases are isolated
systems whose coherence times can be much longer than typical time scales of condensed
matter systems.

\begin{figure}[t]
\centering
\includegraphics[width=0.8\linewidth]{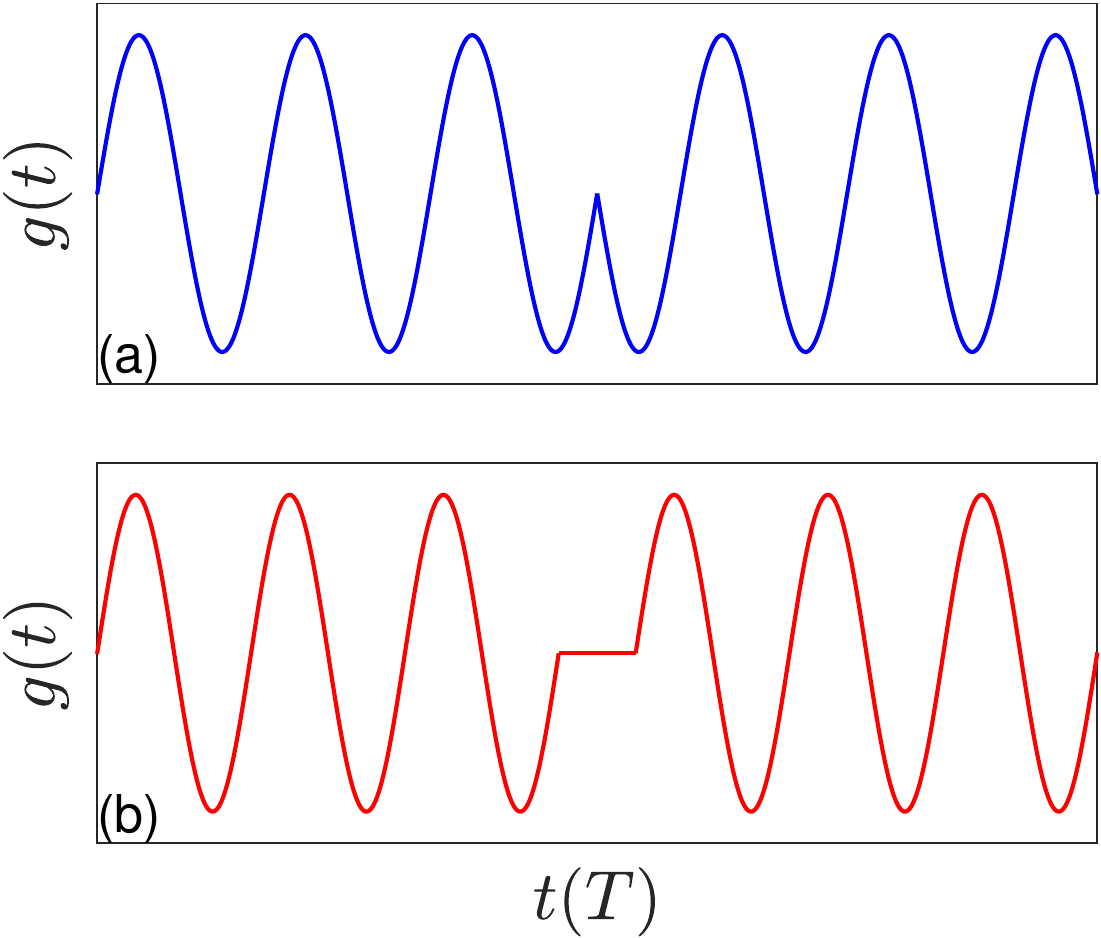}
\caption{The time dependence of the interaction strength. (a) is the
protocol used in the Chicago experiment and (b) is the protocol proposed in
this letter. We will discuss and compare how these two schemes reverse
the quantum evolution of a many-body system. }
\label{scheme}
\end{figure}

One type of dynamics that has been widely explored in ultracold atoms is that under periodical driving~\cite{driving2017}. For instance, the periodical modulation of optical lattices
has been employed to to create artificial magnetic fields~\cite{magnetic_Sengstock2011,magnetic_Bloch2011,magnetic_Sengstock2012,
magnetic_Sengstock2013,magnetic_Bloch2013,magnetic_ketterle2013,magnetic_Bloch2014,magnetic_ketterle2015,magnetic_Greiner2016} and topological bands~\cite{topo_Oka2009,topo_Kitagawa2011,topo_Galitzki2011,topo_Cayssol2013,Zheng,ETH,topo_cooper2014,Hamburg,Aidelsburger,Flaschner} and to realise gauge field with dynamics~\cite{Chicago,
ETH_gauge, Munich_gauge}. Recently the Chicago group has explored the periodical
modulation of the interacting strength between atoms in a weakly interacting
Bose-Einstein condensate confined in a cylindrical box potential \cite
{Chicago_1,Chicago_2}. Such a modulation induces a parametric resonance and leads to an
exponential growth of quasi-particles with energy close to half the modulation frequency~\cite{Wu}. To show that
this many-body dynamics is indeed coherent, in a latest experiment they also
attempted to reverse the many-body dynamics by inverting the time-dependence of the
interaction modulation~\cite{Chicago_3}. To be more precise, the following
time-dependent interaction strength $g(t)$ was considered
\begin{equation}
g(t)=\left\lbrace%
\begin{array}{lc}
g_{0}\sin(\omega t) & 0\leq t\leq n T \\
g_{0}\sin(\omega t-\pi) & n T\leq t\leq 2nT,%
\end{array}%
\right.  \label{case_i}
\end{equation}
where $g_0$ is the oscillation amplitude, $\omega$ is the oscillation
frequency, $T=2\pi/\omega$ is the period and $n$ is an integer. This oscillation scheme is denoted by
protocol (a) and shown in Fig.~1(a). During the first $n$ periods of oscillations $%
0\leq t\leq n T$, atoms are resonantly excited to states whose energy are in the vicinity of $\hbar \omega/2$. In the second $n$ periods of oscillations $n T\leq t\leq
2nT $, however,  a significant portion of those excitations are found to return to the
condensate mode. This is a strong evidence that a coherent many-body
dynamical process can indeed be reversed.

In this letter we present a different oscillation scheme, denoted as the
protocol (b) and shown in Fig.~1(b), which we show can reverse the many-body dynamics to a greater degree than the protocol (a). This scheme is
mathematically described by
\begin{align}
g(t)=\left\lbrace%
\begin{array}{lc}
g_{0}\sin(\omega t) & 0\leq t\leq n T \\
0 & n T\leq t\leq (n+\frac{1}{2})T \\
g_{0}\sin(\omega t-\pi) & (n+\frac{1}{2})T\leq t\leq (2n+\frac{1}{2})T.%
\end{array}%
\right.  \label{case_b}
\end{align}
In this scheme, the driving takes a half period break after the first $n$
periods of oscillation, whereby the system undergoes free evolution governed by the
non-interacting Hamiltonian. The second $n$ periods of oscillation is a repetition of the first,
which can be seen by letting $%
t^\prime=t-T/2$ and writting $g(t^\prime)=g_0\sin(\omega t^\prime)$ for $nT\leq
t^\prime\leq 2nT$. Without the half period pause inserted in between, we would simply have a
single oscillation throughout the entire process and
the quasi-particles will be continuously excited. Thus, the fact that our scheme can reverse the
quasi-particle excitation process is quite counter-intuitive at first
glance.

As we will explain in detail later, the underlying principle by which the protocol (b) reverses the dynamics is reminiscent of the spin echo \cite{Spinecho1,Spinecho2}. Spin echo in a NMR system is a scheme to refocuse the magnetisation against the dephasing due to the inhomogeneous magnetic field. There, the magnetic field under which the spins process
does not change, similar to the fact that our protocol (b) involves exactly the same modulation $g(t)$ in the first and the second $n$ periods of driving. The key of the spin echo effect is a $\pi$ pulse during the spin procession that inverts the spin orientation. In our protocol (b), the
analogy of the $\pi$ pulse is the free evolution that introduces a phase to
the wave function. Because of this close analogy with the spin echo and the many-body nature of our problem, we
refer to the dynamics under our protocol as the ``many-body echo". Although we introduce the concept of
many-body echo using a weakly interacting Bose condensate as an example, the
idea can be generally applied to other many-body systems.

\begin{figure}[t]
\centering
\includegraphics[width=0.9\linewidth]{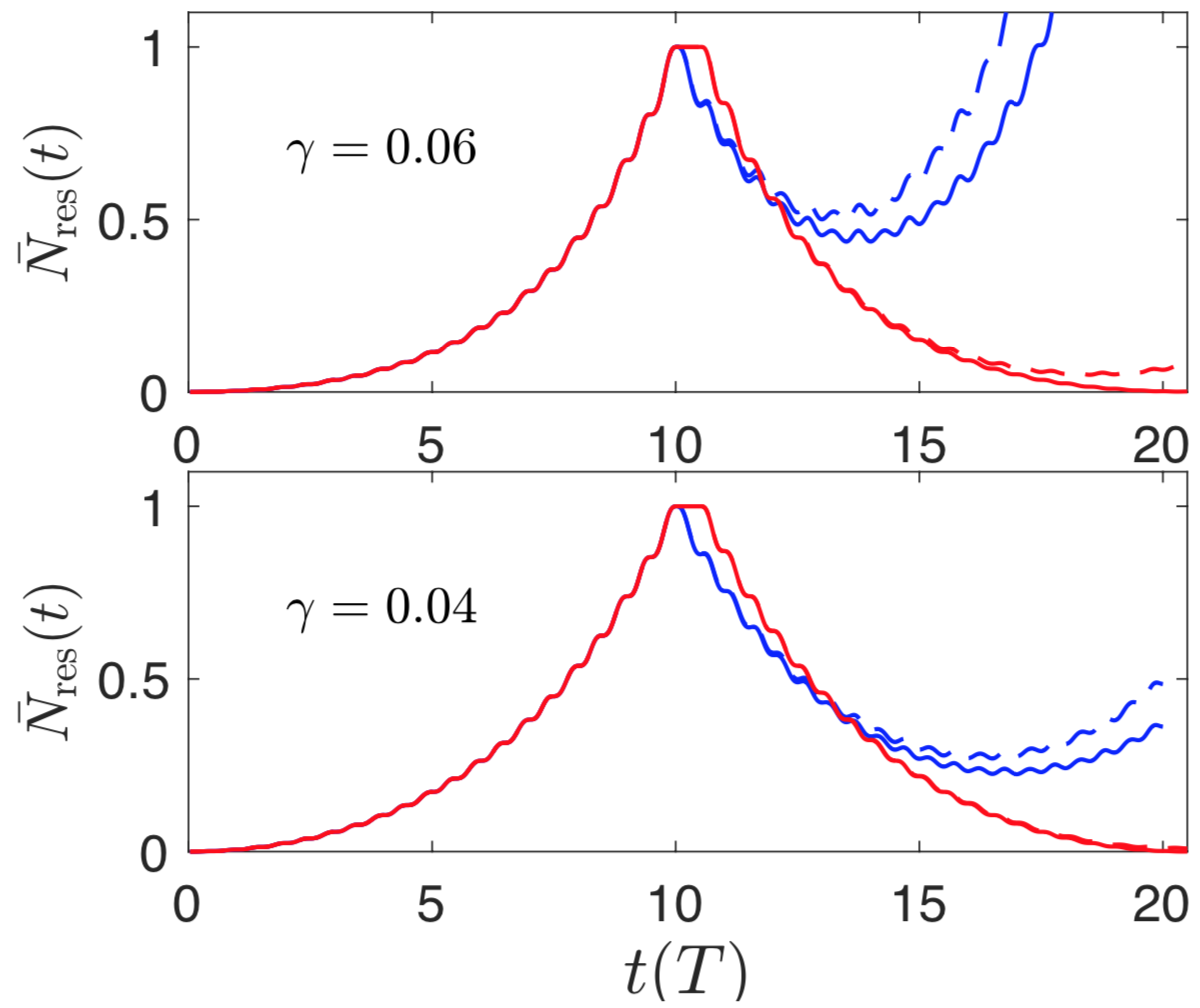}
\caption{Normalised population of resonant excitations as 
a function of time from both protocol (a) (blue) and protocol (b) (red)
depicted in Fig.~\ref{scheme}. 
The dashed lines are the populations at $\epsilon_{\bf k} = \hbar\omega/2$ and the 
the solid lines are those at $\epsilon_{\bf k} = (1/2+3\gamma^2/8)\hbar\omega$.
Calculations are done for two values of modulation strength, i.e., $\protect\gamma%
=0.06$ in the upper panel and $\gamma=0.04$ in the lower panel.}
\label{uniform}
\end{figure}

\textit{Bogoliubov Theory.} We consider a Bose gas with a periodically
modulated interaction, described by the Hamiltonian
\begin{align}  \label{eq:H3d}
\hat{H}=&\int d\mathbf{r}\hat{\psi}^{\dagger}(\mathbf{r}) \left [-\frac{%
\hbar^2 \nabla^2}{2m} + V_{\mathrm{tr}}(\mathbf{r}) \right ] \hat{\psi}(%
\mathbf{r})  \notag \\
&+\frac{g(t)}{2}\int d\mathbf{r}\hat{\psi}^{\dagger}(\mathbf{r})\hat{\psi}%
^{\dagger}(\mathbf{r})\hat{\psi}(\mathbf{r})\hat{\psi}(\mathbf{r}),
\end{align}
where $\hat{\psi}(\mathbf{r})$ is the bosonic field, $V_{\mathrm{tr}}(%
\mathbf{r})$ is the trapping potential, $m$ is the atom mass and $g(t)$ is
the interaction strength. At $t=0$, the system is non-interacting and all
the particles are condensed in the ground state $\varphi_0(\mathbf{r})$ of
the single particle Hamiltonian $\hat h(\mathbf{r}) = -\hbar^2 \nabla^2/(2m)
+ V_{\mathrm{tr}}(\mathbf{r})$. After the interaction modulation is turned
on, we monitor the dynamics by calculating the population on the
 single particle excited state
$\varphi_j(\mathbf{r})$, where $\hat h(\mathbf{r}) \varphi_j(\mathbf{r}) =
\epsilon_j \varphi_j(\mathbf{r}) $. We focus on a regime where the
depletion of the condensate during the interaction modulation is
sufficiently small, such that the dynamics can be well captured by the
following time-dependent Bogliubov-de Gennes(BdG) equations~\cite{Castin}
\begin{align}  \label{bdgeq1}
i\hbar{\partial_t u_j(\mathbf{r},t)} &= \mathcal{L}(\mathbf{r},t) u_j(\mathbf{r}%
,t) - g(t) \Phi^2_0 v_j(\mathbf{r},t) \\
i\hbar{\partial_t v_j(\mathbf{r},t)} &= - \mathcal{L}(\mathbf{r},t) v_j(\mathbf{r}%
,t) + g(t) \Phi_0^{*2} u_j(\mathbf{r},t), \label{bdgeq2}
\end{align}
where the Bogoliubov amplitudes satisfy the orthonormal relations
$
\int d{\mathbf{r}} \left [ u_i(\mathbf{r},t)u^*_j(\mathbf{r},t) - v_i(%
\mathbf{r},t)v^*_j(\mathbf{r},t) \right ] = \delta_{ij}
$
with the initial conditions $u_i(\mathbf{r},0) = \varphi_i(\mathbf{r})$ and $%
v_i(\mathbf{r},0) = 0$. Here
$
\mathcal{L} (\mathbf{r}, t) \equiv \hat h(\mathbf{r})+2g(t)|\Phi_0(\mathbf{r}%
,t)|^2 - \mu
$,
where $\mu$ is the initial chemical potential and $\Phi_0(\mathbf{r},t) =
\sqrt{N_0(t)}\varphi_0(\mathbf{r})$ is the time-dependent condensate wave
function. The number of particles excited to the state $\varphi_j(\mathbf{r})$ is given by
$
N_j(t) = \int d {\mathbf{r}} |v_j(\mathbf{r},t)|^2
$.
The BdG equations are solved together with number conservation condition
$N = N_0(t) + \sum_j N_j(t)
$.

To illustrate the essential physics involved, we first consider a uniform
condensate and, for simplicity, take the condensate
density $|\Phi_0(\mathbf{r}, t)|^2$ to be a constant $n_0$ independent of time. This approximation
is not necessary for the numerical calculation but will simplify our later analysis without compromising the main results.
 As mentioned earlier, a periodical modulation of the interaction with frequency $\omega$ mostly excites the atoms to
  states with energy $\epsilon_{\mathbf k}\equiv \hbar^2\mathbf{k}^2/(2m) \sim \hbar\omega/2$, because two atoms with
  opposite momentum collide and absorb one quanta of energy $\hbar\omega$.
Shown in Fig.~\ref{uniform} are the population of atoms excited to the resonant energy  $\epsilon_{\mathbf k} = \hbar\omega/2$ and to a slightly modified resonant energy $\epsilon_{\mathbf k} = (1/2+3\gamma^2/8)\hbar\omega$ (the significance of this modification will be explained later), calculated for the interaction modulations depicted in both protocol (a) and (b). Here $\gamma = {g_0 n_0}/{(\hbar \omega)}
$ is a relatively small, dimensionless parameter that characterises the strength of the modulation.
As we can see, for both schemes, the atoms are
excited during the first stage of interaction modulation, but most of them
are absorbed back to condensate after the second stage. It is also clear that the
protocol (b) reverses the many-body process much better than the protocol (a),
particularly for larger modulation strengths.

\textit{Floquet Hamiltonian.} We first present our understanding of the
above phenomenon in terms of the Floquet Hamiltonian, which governs the
stroboscopic evolution of the system. We begin with the Bogoliubov
Hamiltonian for the uniform condensate  $\hat{H}_{%
\text{Bg}}(t)=g(t)N^2/(2V)+\sum_{\mathbf{k}}\hat{H}^{\mathbf{k}}_{\text{Bg}}(t)$, where $V$ is the volume and
\begin{equation}
\hat{H}^{\mathbf{k}}_{\text{Bg}}(t)=\left[\epsilon_{\mathbf{k}}+g(t)n_0
\right]\hat{a}^{\dagger}_{\mathbf{k}}\hat{a}_{\mathbf{k}} +\frac{n_0g(t)}{2}%
\left(\hat{a}^\dag_{\mathbf{k}}\hat{a}^\dag_{\mathbf{-k}}+\text{h.c.}
\right). \label{Bogoliubov}
\end{equation}
To derive the Floquet Hamiltonian using the high frequency expansion, it is necessary to
first apply the rotating frame transformation
\begin{equation}
\hat{R}(t)=\exp \left( \frac{i\omega t}{2}\sum\limits_{\mathbf{k}\neq 0}\hat{%
a}_{\mathbf{k}}^{\dag }\hat{a}_{\mathbf{k}}\right) \label{Rt}
\end{equation}%
to eliminate the resonance energy term. In doing so,
the Hamiltonian in the rotating frame is given by $\hat{H}_{\text{R}}(t)=%
\hat{R}(t)[\hat{H}_{\text{Bg}}(t)-i\hbar \partial _{t}]\hat{R}^{\dag }(t)$,
which yields $\hat{H}_{R}(t)=g(t)N^2/(2V)+\sum_{\mathbf{k}}\hat{H}_{R}^{\mathbf{k}}(t)$ with
\begin{align}
\hat{H}_{R}^{\mathbf{k}}(t)=& \left[ \epsilon _{\mathbf{k}}-\frac{\hbar
\omega }{2}+g(t)n_{0}\right] \hat{a}_{\mathbf{k}}^{\dag }\hat{a}_{\mathbf{k}}
\notag  \label{eq:H_R} \\
& +\frac{g(t)n_{0}}{2}\left( e^{i\omega t}\hat{a}_{\mathbf{k}}^{\dag }\hat{a}%
_{\mathbf{-k}}^{\dag }+e^{-i\omega t}\hat{a}_{\mathbf{k}}\hat{a}_{\mathbf{-k}%
}\right) .
\end{align}%
For an interaction strength $g(t)$ periodical in $T$, an effective Floquet
Hamiltonian $\hat{H}_{\mathrm{eff}}$ capturing the evolution at integer periods of oscillation can be introduced by
\begin{equation}
\mathcal{T}\exp \left( -\frac{i}{\hbar}\int_{\alpha T}^{\alpha T+T}\hat{H}_{R}(t)dt\right)
=\exp \left( -\frac{i}{\hbar}\hat{H}_{\mathrm{eff}}T\right),  \label{eq:Floquet}
\end{equation}%
where $\mathcal{T}$ is the time-ordering operator and $\alpha$ specifies
the initial reference time. By Fourier transforming $\hat{H%
}_{R}(t)=\sum_{p}\hat{H}_{p}e^{ip\omega t}$ and using the $1/\omega $
expansion, we obtain~\cite{Maricq}
\begin{align}
& \hat{H}_{\mathrm{eff}}\approx \hat{H}_{0}+  \notag
\label{eq:H_eff_general} \\
& \sum_{p>0}\left( \frac{\left[ \hat{H}_{p},\hat{H}_{-p}\right] }{p\hbar
\omega }-\frac{\left[ \hat{H}_{p},\hat{H}_{0}\right] }{p\hbar \omega
e^{-i2p\alpha \pi }}+\frac{\left[ \hat{H}_{-p},\hat{H}_{0}\right] }{p\hbar
\omega e^{i2p\alpha \pi }}\right).
\end{align}%
The effective Floquet Hamiltonian we define here is different from the
conventional one \cite{Heff2014,Heff2015}, in which the information of the
initial state is absorbed in the kick operator.

For $0\leq t\leq nT$, the time dependence of $g(t)$ is the
same for protocol (a) and (b). Writing $\hat{H}_{\mathrm{eff}} = \sum_{\mathbf{k}}\hat{H}_{\mathrm{eff}}^{\mathbf{k}}$ and following Eq.~\ref%
{eq:H_eff_general}, we obtain for this duration
\begin{align}
\label{eq:Heff_1}
\frac{\hat{H}_{\mathrm{eff}}^{\mathbf{k}}}{\hbar \omega }=-\frac{1}{2}\gamma
\hat{A}_{y}^{\mathbf{k}}-{\gamma ^{2}}\hat{A}_{x}^{\mathbf{k}}+\Delta _{%
\mathbf{k}}\left( 2\hat{A}_{z}^{\mathbf{k}}-1\right) ,
\end{align}%
where $\Delta _{\mathbf{k}%
}\equiv \epsilon_{\bf k}/\hbar \omega -1/2-{3\gamma ^{2}}/{8}$ and
\begin{eqnarray*}
\hat{A}_{x}^{\mathbf{k}} &\equiv &\frac{1}{2}\left( \hat{a}_{\mathbf{k}%
}^{\dag }\hat{a}_{\mathbf{-k}}^{\dag }+\hat{a}_{\mathbf{k}}\hat{a}_{\mathbf{%
-k}}\right) , \\
\hat{A}_{y}^{\mathbf{k}} &\equiv &\frac{1}{2i}\left( \hat{a}_{\mathbf{k}%
}^{\dag }\hat{a}_{\mathbf{-k}}^{\dag }-\hat{a}_{\mathbf{k}}\hat{a}_{\mathbf{%
-k}}\right) , \\
\hat{A}_{z}^{\mathbf{k}} &\equiv &\frac{1}{2}\left( \hat{a}_{\mathbf{k}%
}^{\dag }\hat{a}_{\mathbf{k}}+\hat{a}_{\mathbf{-k}}\hat{a}^{\dag }_{\mathbf{%
-k}}\right) .
\end{eqnarray*}%
These three operators form the group of pseudo-rotations, SU(1,1), obeying
the commutation relations $\left[ \hat{A}_{x}^{\mathbf{k}},\hat{A}_{y}^{%
\mathbf{k}}\right] =-i\hat{A}_{z}^{\mathbf{k}}$, $\left[ \hat{A}_{y}^{%
\mathbf{k}},\hat{A}_{z}^{\mathbf{k}}\right] =i\hat{A}_{x}^{\mathbf{k}}$, and
$\left[ \hat{A}_{z}^{\mathbf{k}},\hat{A}_{x}^{\mathbf{k}}\right] =i\hat{A}%
_{y}^{\mathbf{k}}$.

The second $n$ periods of oscillation in the protocol (a) and (b) are governed by
 different Floquet Hamiltonians. For protocol (a),  we find the following effective Hamiltonian
\begin{align}
\frac{\hat{H}_{\mathrm{eff,a}}^{\mathbf{k}}}{\hbar \omega }=\frac{1}{2}%
\gamma \hat{A}_{y}^{\mathbf{k}}-{\gamma ^{2}}\hat{A}_{x}^{\mathbf{k}}+\Delta
_{\mathbf{k}}\left( 2\hat{A}_{z}^{\mathbf{k}}-1\right)
\end{align}%
for $nT\leq t\leq 2nT$. If we consider the resonant modes  $\epsilon _{\mathbf{k}}\sim\hbar
\omega /2$ such that $\Delta_{\mathbf{k}} \sim 3\gamma^2/8 $, it is clear that $\hat{H}_{\text{eff,a}}^{\mathbf{k}}$
inverts $\hat{H}_{\text{eff}}^{\mathbf{k}}$ up to the leading order of $%
\gamma $, but not to the second order of $\gamma ^{2}$.

Now consider the protocol (b). During the half period
of free evolution $nT\leq t\leq (n+1/2)T$, the Hamiltonian in the rotating frame vanishes
for the resonant modes with $\epsilon _{\mathbf{k}%
}\sim \hbar \omega /2$.  For
$nT+T/2\leq t\leq 2nT+T/2$, even though the functional form of $g(t)$ is the same as that in the
second stage of protocol (a), the initial reference time characterized by parameter $\alpha $
in Eq.~\ref{eq:Floquet}  is different. More specifically, we have $\alpha =0$ for the protocol (a) and $\alpha =1/2$ for
the protocol (b). In the Floquet theory, this results in a difference in
the so-called micro-motion term in the effective Hamiltonian~\cite{Heff2014}. Thus we find the effective Hamiltonian of the
protocol (b) as
\begin{align}
\frac{\hat{H}_{\mathrm{eff,b}}^{\mathbf{k}}}{\hbar \omega }=\frac{1}{2}%
\gamma \hat{A}_{y}^{\mathbf{k}}+{\gamma ^{2}}\hat{A}_{x}^{\mathbf{k}}+\Delta
_{\mathbf{k}}\left( 2\hat{A}_{z}^{\mathbf{k}}-1\right)
\end{align}%
for $nT+T/2\leq t\leq 2nT+T/2$.
Now we can see that both the first and the second term in $\hat H_{\mathrm{eff,b}%
}^{\mathbf{k}}$ are opposite to those in $\hat H_{\mathrm{eff}}^{\mathbf{k}}$. If
we further consider the resonance modes specified by $\Delta _{\mathbf{k}}=0$,
i.e., $\epsilon_{\bf k} = (1/2+3\gamma^2/8)\hbar\omega$, $%
\hat H_{\mathrm{eff,b}}^{\mathbf{k}}$ completely inverts $\hat H_{\mathrm{eff}%
}^{\mathbf{k}}$ for all contributions up to the order of $\gamma ^{2}$. This
explains why the protocol (b) reverses the many-body dynamical process better than the
protocol (a). Since the difference between the two protocols lies in the second order terms of $\gamma$ in
the effective Hamiltonian, it
also explains why the difference is more significant for larger $\gamma $,
as shown in Fig.~\ref{uniform}. Finally, it can be shown that all the excitations with $\Delta_{\bf k} \ll \gamma$
will be well reversed by our protocol.

\begin{figure}[t]
\centering
\includegraphics[width=0.9\linewidth]{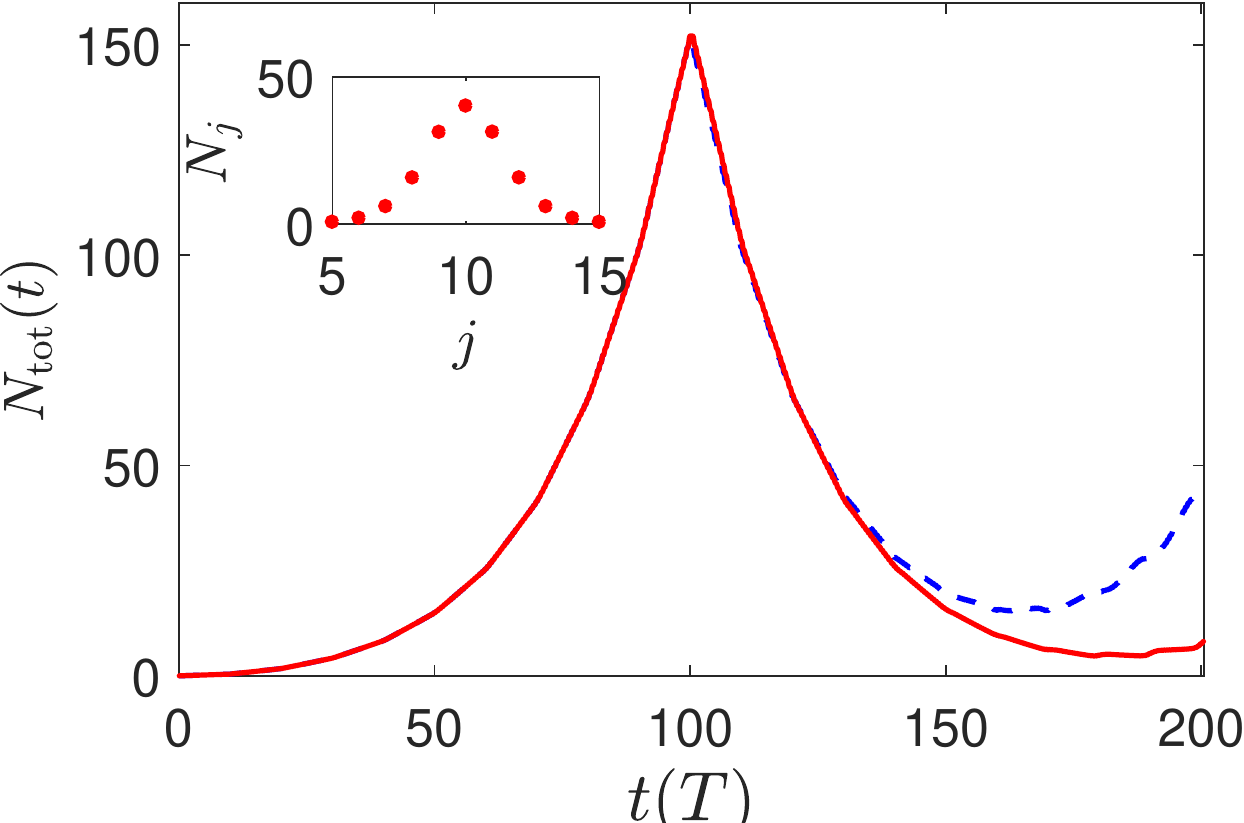}
\caption{The total number of excitations as a function of time  in a quasi-one-dimensional harmonic trapped  system.
 The dashed (blue) line and the solid (red) line correspond to the protocol
(a) and the protocol (b) respectively. The inset shows that the
occupation of different single-particle modes at $t=100T$ when the number of
excitation reaches its maximum. }
\label{trap}
\end{figure}

\textit{Many-Body Echo.} Now we discuss the connection between the
underlying physics of the protocol (b) and the spin echo. For this purpose we
introduce an alternative approach to understand the reversal of dynamics.
As mentioned earlier, the first and second $n$ periods of oscillation in the protocol (b) are identical, which
 can be seen by writing $t^{\prime }=t-T/2$,  such that $g(t^{\prime })=g_{0}\sin
(\omega t^{\prime })$ for $nT\leq t^{\prime }\leq 2nT$. However, they become
inequivalent when viewed from the single rotating frame of reference introduced, leading to different Floquet Hamiltonians obtained earlier.
Such an equivalence can be restored if we apply the unitary rotation $\hat{R}(t)$ for $0\leq t\leq nT$ and
another $\hat{R}(t^{\prime })$ for $nT\leq t^{\prime }\leq 2nT
$ (i.e. $nT+T/2\leq t\leq 2nT+T/2$).  In
this approach, the free evolution for the intermediate half period $nT\leq t\leq nT+T/2$ is according to the
original Bogoliubov Hamiltonian in Eq.~\ref{Bogoliubov}, but the system will be governed by the same effective Hamiltonian $\hat{H}_{\text{eff}}$
in Eq.~\ref{eq:Heff_1} during both sections of the driving. Hence,
the total evolution operator from $t=0$ to $%
t=2nT+T/2$ is given by%
\begin{eqnarray}
\hat{U} &=&\hat{R}^{\dag }(2nT)e^{-\frac{i}{\hbar}\hat{H}_{\text{eff}}nT}\hat{R}(nT)  \notag
\\
&&\times e^{-i\frac{T}{2\hbar}\sum_{\bf k}\epsilon _{\mathbf{k}}\hat{a}_{\mathbf{k}}^{\dag }%
\hat{a}_{\mathbf{k}}}%
\hat{R}^{\dag }(nT)e^{-\frac{i}{\hbar}\hat{H}_{\text{eff}}nT}\hat{R}(0).
\end{eqnarray}%
Restricting ourselves to resonance modes with $\epsilon _{\mathbf{k}}\sim \hbar \omega /2$
and using $\hat{R}(nT)=(-1)^n$, we obtain $\hat U = \prod_{\bf k}\hat U_{\bf k}$ where
\begin{equation}
\hat{U}_{\bf k}=e^{-\frac{i}{\hbar}\hat{H}^{\bf k}_{\text{eff}}nT}e^{-i\pi \hat{A}_{z}^{%
\mathbf{k}}}e^{-\frac{i}{\hbar}\hat{H}^{\bf k}_{\text{eff}}nT}.
\end{equation}%
The operator $e^{-i\pi \hat{A}_{z}^{\mathbf{k}}}$ is reminiscent of the $\pi $-pulse inserted in the spin echo experiment.
More precisely, this operator acts on the Bogoliubov-type many-body ground state 
as%
\begin{align}
e^{-i\pi \hat{A}_{z}^{\mathbf{k}}}e^{%
\left ( \chi_{{\bf k}} \hat{a}_{\mathbf{k}%
}^{\dag }\hat{a}_{\mathbf{-k}}^{\dag }-\chi ^{\ast }_{{\bf k}}\hat{a}_{%
\mathbf{k}}\hat{a}_{\mathbf{-k}}\right ) }\left\vert 0\right\rangle   =e^{-\left ( \chi_{{\bf k}} \hat{a}_{%
\mathbf{k}}^{\dag }\hat{a}_{\mathbf{-k}}^{\dag }-\chi ^{\ast }_{{\bf k}}%
\hat{a}_{\mathbf{k}}\hat{a}_{\mathbf{-k}}\right ) }\left\vert 0 \right\rangle.\nonumber
\end{align} We note that this operator adds a $\pi$ phase shift to the wave function of excitations,
which plays a key role in reversing the many-body dynamics. 

\textit{Harmonic Trapped Case.} For the uniform system, the resonance of
excitations due to the interaction modulation has a typical width of
the order of $\gamma \hbar\omega$, while our
protocol only reverses those satisfying $|\epsilon_{\bf k}-\hbar\omega/2 |\ll \gamma \hbar\omega$.
In order to achieve a complete reversal of all excitations,
we turn to a quasi-one-dimensional Bose condensate in a harmonic trap with the
frequency $\omega_z$. The advantage of this setup is that the single particle eigen-energy
$\epsilon_j=(j+1/2)\hbar\omega_z$ is discrete, such that only pairs of particles with $%
(j_1+j_2)\omega_z=\omega$ can be excited if the level separation $%
\hbar\omega_z$ is large than or comparable with the amplitude of interaction energy
modulation. In such a case, almost all the excitations can be reversed by our protocol.

To demonstrate this, we consider a condensate with a strong transverse
confinement $\omega_\perp=2\pi \times 430$Hz such that the modulation will
only excite the axial modes. The condensate thus behaves like a one-dimensional system
with an effective interaction modulation amplitude $\tilde g_0 = g_0/(2\pi l_\perp^2)$, where
$l_\perp = \sqrt{\hbar/m\omega_\perp}$. The modulation frequency is chosen to be
$\omega = 20\omega_z$, where the axial trapping frequency is $\omega_z=2\pi \times 200$Hz.
We numerically solve the number-conserving BdG equations described
earlier for this system with a total atom number $N=730$ and a modulation strength
 $\tilde\gamma = \frac{\tilde g_0 \tilde n_0}{\hbar\omega }=0.12$, where $\tilde n_0 = N/l_z$ with $l_z = \sqrt{\hbar/m\omega_z}$. Shown in Fig.~\ref{trap}
are the total number of excitations $N_\text{tot}(t)=\sum_j N_j(t)$ from both the protocol (a) and (b).
As shown in the inset of Fig.~\ref{trap}, the occupied modes indeed mostly satisfy the
resonance condition $j_1+j_2=\omega/\omega_z$. We see that our protocol, again much more effective than the protocol (a), achieves an almost perfect reversal of
all the excitations.

\textit{Outlook.} In summary, we have developed an analogy of the spin echo
in a Floquet quantum many-body system, which we refer to as the many-body echo.
Although we demonstrate our protocol to realise the many-body echo in a weakly interacting Bose gas, we believe
this method can be generalised to other quantum many-body systems under periodical
driving and will find broad applications in future
research of Floquet quantum matter. One application, for instance, could be
facilitating the experimental measurement of the out-of-time-ordered
correlation function by reversing the many-body
dynamics ~\cite{OTOC1,OTOC2,OTOC3}. Another application draws inspiration from the spin echo effect, where
the imperfect refocusing can be used to detect decoherence time due to spin-spin
interactions. Similarly, in a many-body system with significant quasi-particles interactions,
the degree to which
 our protocol does not reverse the dynamics can then be attributed to the quasi-particle interactions. Finally, concerning a generic issue of Floquet driving, the system will eventually be heated to infinite temperature as it keeps absorbing energy from the driving \cite{Floquet_Heating2014}. Thus one has to reply on effects such as the many-body localization
to prevent heating and allows the formation of interesting phases in the Floquet
system~\cite{Yao}. Our protocol opens up an alternative route for preventing heating
whereby novel Floquet physics may become accessible.

\textit{Acknowledgement}. This work is supported by Beijing Distinguished Young Scientist Program (HZ), MOST under Grant No. 2016YFA0301600 (HZ) and NSFC Grant No. 11734010 (HZ).


\begin{thebibliography}{99}
\bibitem{chaos1} E. Altman, Nat. Phys. {\bf 14}, 979 (2018).

\bibitem{chaos2} X.-L. Qi, Nat. Phys. {\bf 14}, 984 (2018).

\bibitem{chaos3} B. Swingle, Nat. Phys. {\bf 14}, 988 (2018).

\bibitem{driving2017} A. Eckardt, Rev. Mod. Phys. \textbf{89}, 011004 (2017).



\bibitem{magnetic_Sengstock2011}
J. Struck, C. \"{O}lschl\"{a}ger, R. Le Targat, P. Soltan-Panahi, A. Eckardt,
M. Lewenstein, P. Windpassinger, and K. Sengstock, Science \textbf{333}, 996 (2011).

\bibitem{magnetic_Bloch2011}
M. Aidelsburger, M. Atala, S. Nascimb¨¨ne, S. Trotzky, Y.-A. Chen, and I. Bloch,  Phys. Rev. Lett. \textbf{107}, 255301 (2011).


\bibitem{magnetic_Sengstock2012}
J.Struck, C. \"{O}lschl\"{a}ger, M. Weinberg, P. Hauke, J. Simonet, A. Eckardt, M. Lewenstein, K. Sengstock, and P. Windpassinger, Phys. Rev. Lett. \textbf{108}, 225304 (2012)

\bibitem{magnetic_Sengstock2013}
J. Struck, M. Weinberg, C. \"{O}lschl\"{a}ger, P. Windpassinger, J. Simonet, K. Sengstock, R. H\"{o}ppner, P. Hauke, A. Eckardt, M. Lewenstein and L. Mathey, Nat. Phys. \textbf{9}, 738 (2013)

\bibitem{magnetic_Bloch2013}
M. Aidelsburger, M. Atala, M. Lohse, J. T. Barreiro, B. Paredes, and I. Bloch, Phys. Rev. Lett. \textbf{111}, 185301 (2013). 

\bibitem{magnetic_ketterle2013}
H. Miyake, G. A. Siviloglou, J. Kennedy, W. C. Burton, and W. Ketterle, Phys. Rev. Lett. \textbf{111}, 185302 (2013)

\bibitem{magnetic_Bloch2014}
M. Atala, M. Aidelsburger, M. Lohse, J. T. Barreiro, B. Paredes, and I. Bloch, Nat. Phys. \textbf{10}, 588. (2014)

\bibitem{magnetic_ketterle2015}
C. J. Kennedy, W. C. Burton, W. C. Chung, and W. Ketterle, Nat. Phys. \textbf{11}, 859 (2015)

\bibitem{magnetic_Greiner2016}
M. E. Tai, A. Lukin, M. Rispoli, R. Schittko, T. Menke, D. Borgnia, P. M. Preiss, F. Grusdt, A. M. Kaufman, and Markus Greiner, Nature \textbf{546}, 519 (2017)

\bibitem{topo_Oka2009}
T. Oka, and H. Aoki, Phys. Rev. B \textbf{79}, 081406 (2009)

\bibitem{topo_Kitagawa2011}
T. Kitagawa, T. Oka, A. Brataas, L. Fu, and E. Demler, Phys. Rev. B \textbf{84}, 235108 (2011)

\bibitem{topo_Galitzki2011}
N. H. Lindner, G. Refael, and V. Galitzki, Nat. Phys. \textbf{7}, 490 (2011)

\bibitem{topo_Cayssol2013}
J. Cayssol, B. D\'{o}ra, F. Simon, and R. Moessner, Phys. Status Solidi RRL \textbf{7}, 101. (2013). 

\bibitem{Zheng} W. Zheng and H. Zhai, Phys. Rev. A \textbf{89}, 061603(R)
(2014).

\bibitem{ETH} G. Jotzu, M. Messer, R. Desbuquois, M. Lebrat, T. Uehlinger,
D. Greif and T. Esslinger, Nature \textbf{515}, 237 (2014).

\bibitem{topo_cooper2014}
S. K. Baur, M. H. Schleier-Smith, and N. R. Cooper, Phys. Rev. A \textbf{89}, 051605(R) (2014)

\bibitem{Aidelsburger} M. Aidelsburger, M. Lohse, C. Schweizer, M. Atala, J. T. Bar-reiro,
S. Nascimb\'{e}ne, N. R. Cooper, I. Bloch, and N. Goldman, Nat. Phys. {\bf 11}, 162 (2015).

\bibitem{Flaschner} N. Fl\"{a}schner,  B. S. Rem,  M. Tarnowski,  D. Vogel,  D.-S. L\"{u}h-mann,  K.  Sengstock,  and  C.  Weitenberg,
Science {\bf 352}, 1091 (2016).

\bibitem{Hamburg} M. Tarnowski, F. Nur\"{U}nal, N. Fl\"{a}schner, B. S. Rem,
A. Eckardt, K. Sengstock and C. Weitenberg, Nat. Commun. \textbf{10}, 1728
(2019).

\bibitem{Chicago} L. W. Clark, B. M. Anderson, L. Feng, A. Gaj, K. Levin, C.
Chin, Phys. Rev. Lett. \textbf{121}, 030402 (2018)

\bibitem{ETH_gauge} F. G\"{o}rg, K. Sandholzer, J. Minguzzi, R. Desbuquois,
M. Messer, and T. Esslinger, Nat. Phys. (2019), 10.1038/s41567-019-0615-4.

\bibitem{Munich_gauge} C. Schweizer, F. Grusdt, M. Berngruber, L. Barbiero,
E. Demler, N. Goldman, I. Bloch, and M. Aidelsburger, arXiv: 1901.07103.

\bibitem{Chicago_1} L. W. Clark, A. Gaj, L. Feng, and C. Chin, Nature
\textbf{551}, 356 (2017).

\bibitem{Chicago_2} L. Feng, J. Hu, L. W. Clark, and C. Chin,  Science {\bf 363}, 521 (2019).

\bibitem{Wu} Z. Wu and H. Zhai, Phys. Rev. A \textbf{99}, 063624 (2019).

\bibitem{Chicago_3} J. Hu, L. Feng, Z. Zhang, and C. Chin, Nat. Phys.
\textbf{15}, 785 (2019).

\bibitem{Spinecho1} E. L. Hahn, Phys. Rev. \textbf{80},
580 (1950).

\bibitem{Spinecho2} H. Y. Carr, and E. M. Purcell, Phys. Rev. \textbf{94},
630 (1954).

\bibitem{Castin} Y. Castin and R. Dum, Phys. Rev. Lett. {\bf 79}, 3553 (1997).

\bibitem{Maricq} M. M. Maricq, Phys. Rev. B {\bf 25}, 6622 (1982).

\bibitem{Heff2014} N. Goldman, and J. Dalibard, Phys. Rev. X \textbf{4},
031027 (2014).

\bibitem{Heff2015} N. Goldman, J. Dalibard, M. Aidelsburger, and N. R.
Cooper, Phys. Rev. A \textbf{91}, 033632 (2015).

\bibitem{OTOC1}
B. Swingle, G. Bentsen, M. Schleier-Smith, and P. Hayden, Phys. Rev. A {\bf 94}, 040302 (2016).
\bibitem{OTOC2}
G. Zhu, M. Hafezi, and T. Grover, Phys. Rev. A {\bf 94}, 062329 (2016).

\bibitem{OTOC3}
H. Shen, P. Zhang, R. Fan and H. Zhai, Phys. Rev. B, {\bf 96}, 054503 (2017)

\bibitem{Floquet_Heating2014}
L. D'Alessio and M. Rigol, Phy. Rev. X \textbf{4}, 041048 (2014).

\bibitem{Yao}
N. Y. Yao and C. Nayak, Physics Today {\bf 71}, 9, 40 (2018).

\end{thebibliography}
\end{document}